\documentclass[a4paper,twocolumn,pre]{revtex4-1}
\usepackage{graphicx}
\usepackage{amsmath}           % essential
\usepackage{amsfonts}          % AMS fonts
\usepackage{epstopdf}
\usepackage{amsthm}

\usepackage{newtxtext,newtxmath}

\usepackage{tikz,pgfplots,bbm,bm,mathrsfs,dsfont}
\usepackage{subcaption}
\usepackage{color}
\usepackage{quoting}

\usepackage[english]{babel}
\usepackage{hyperref}
\hypersetup{pdfauthor={Capelli Caracciolo DiGioacchino Malatesta},pdftitle={Exact bound for a TSP cost in two dimensions},%
	colorlinks, linktocpage=true, pdfstartpage=3, pdfstartview=FitV,%
	urlcolor=orange, linkcolor=blue, citecolor=red, %pagecolor=RoyalBlue,%
}

\newcommand{\be}{\begin{equation}}
\newcommand{\ee}{\end{equation}}

\usepackage{dsfont}

\begin{document}

\title{How superlocalization affects  Vibrational Energy Exchange process in proteins}
\author{Luca Maggi}\email{l.maggi@fz-juelich.de}
\affiliation{Computational Biomedicine Section, Institute of Advanced Simulation IAS-5 and Institute of Neuroscience and Medicine INM-9, Forschungszentrum J\"ulich, Wilhelm-Johnen-Stra\ss{}e, 52425 J\"ulich, Germany}

\date{\today}
\begin{abstract}
Recent experimental findings on a protein showed the diffusion of vibrational energy does not occur along the backbone interaction,as it might be expected, but prevalently on non-bonded contacts. These results are explained presenting a theoretical picture, supported by computational calculations, that accounts for these different behaviors in vibrational energy exchange process showing the collective motions on the backbone present a $superlocalized$ nature as their decay with the distance $r$ is $exp(-r^{d})$ with $d \sim 1.8$, whereas those associated to non-bonded contacts result simply localized with $d \sim 1$.
\end{abstract}
\maketitle

The proteins primary structure consists in a sequence of different monomers, called residues. Each of them is connected with the adjacent ones in the sequence via covalent bonds and interacts with all the others  through a broad range of weaker non-bonded chemical interactions modeled, for instance, by means of  Lennard-Jones and electrostatic potentials \cite{finkelstein}. The sequence encodes  the three-dimensional protein structure  \cite{finkelstein}, namely the secondary and the tertiary structure, which we can refer to as  topology. The protein topology results in a combination of  short range ordered (e.g. alpha helices or beta sheets) and disordered parts, which differently arranged in the space gives birth to a complex structure. It does not present long range ordering and shares common features with disordered solids \cite{volk} \cite{ciliberti} and fractals \cite{dewey}\cite{leitner}.  
The disordered space arrangement deeply  affects the protein  internal dynamics from single residue  to protein larger collective motions\cite{zhou}\cite{haddadian}. The latter ensue from the coupling  between a single residue displacement  with a distant one. Therefore, they underlie the exchange of vibrational energy ( $E_{vib} \;xc$) among residues\cite{leitner2} \cite{leitnerb} . The disordered topology together with the wealth of possible different chemical interactions, which are the constitutive elements of our depicted $protein \; system$,  confer to collective motions particular properties. These are reflected in the high peculiarity of $E_{vib}xc$ in proteins which can sometimes  present  counter-intuitive features. 
For instance, recent experimental \cite{kondoh} \cite{yamashita} works showed the vibrational energy does not flows through stronger covalent bonds of   sequence (backbone), as it could be thought reasonably  since they are stiffer  and more prone to transfer any kind of displacement, on the contrary  it diffuses along the weaker interactions (contacts) made by non-adjacent residues. In this work we present a theoretical picture, supported by computational calculations, able to explain these findings. We will show how this experimental result are deeply  connected to topology and in particular with geometrical properties which presents scaling rules featuring fractal structure.  
Studying how  $E_{vib}xc$ occurs in protein structures might highly helpful for shedding the light on relevant phenomena strongly associated with protein biological task \cite{leitnerb} as for instance conformational changes\cite{bastida}  or allosteric modulation\cite{li}. 
In order to achieve this goal in this work we defined  a general potential energy for  protein system. We then evaluated the dynamical matrix whose eigenvectors represents the collective motions  and investigated their localization properties. Finally A theoretical picture is presented for accounting the differences in $E_{vib} \;xc$ between backbone and contacts. Our reasoning is supported by computational calculations  carried out , except whereas specified, on 15 different proteins with a sequence length ranging from 100 to 900 residues

The Anisotropic Network Model \cite{tirion} has been employed to reproduce protein collective motions, which has been shown to  be qualitatively accurate for this kind of systems. We have corse-grained  our protein, considering just the center of mass of each protein residue. The considered potential energy is:
$$
\mathcal{U}( \mathbf{r}_{1}....\mathbf{r}_{M}) =\sum_{ij} \mathcal{U}_{ij} ( | \mathbf{r}_{ij}| )=\sum_{ij} \gamma_{ij} (  |\mathbf{r}_{ij}| -|\mathbf{r}_{ij}^{o}| )^{2}.
$$
 Where $\mathbf{r}_{1...M}$ are the co-ordinates of the corresponding residue and $\mathbf{r}_{ij}$ and $\mathbf{r}_{ij}^{o}$   are the distances between the i-th and j-th residue at each time and at the equilibrium respectively. Contributions coming from residues distant more than 15 A were assumed $0$ . $\gamma_{ij}$ is a  constant coupling the i-th and the j-th residue. when $i \neq j$  $\gamma_{ij}$ is taken as  : 
 $$
\gamma_{ij} =
\left\{
\begin{array}{rl}
\gamma_{back}       & \mbox{if } j \pm i \\
\\
\dfrac{\gamma_{back}}{10} &  otherwise
\end{array}
\right.
$$

\begin{figure}
 \includegraphics[scale=0.27,trim={0 4cm 0 0}]{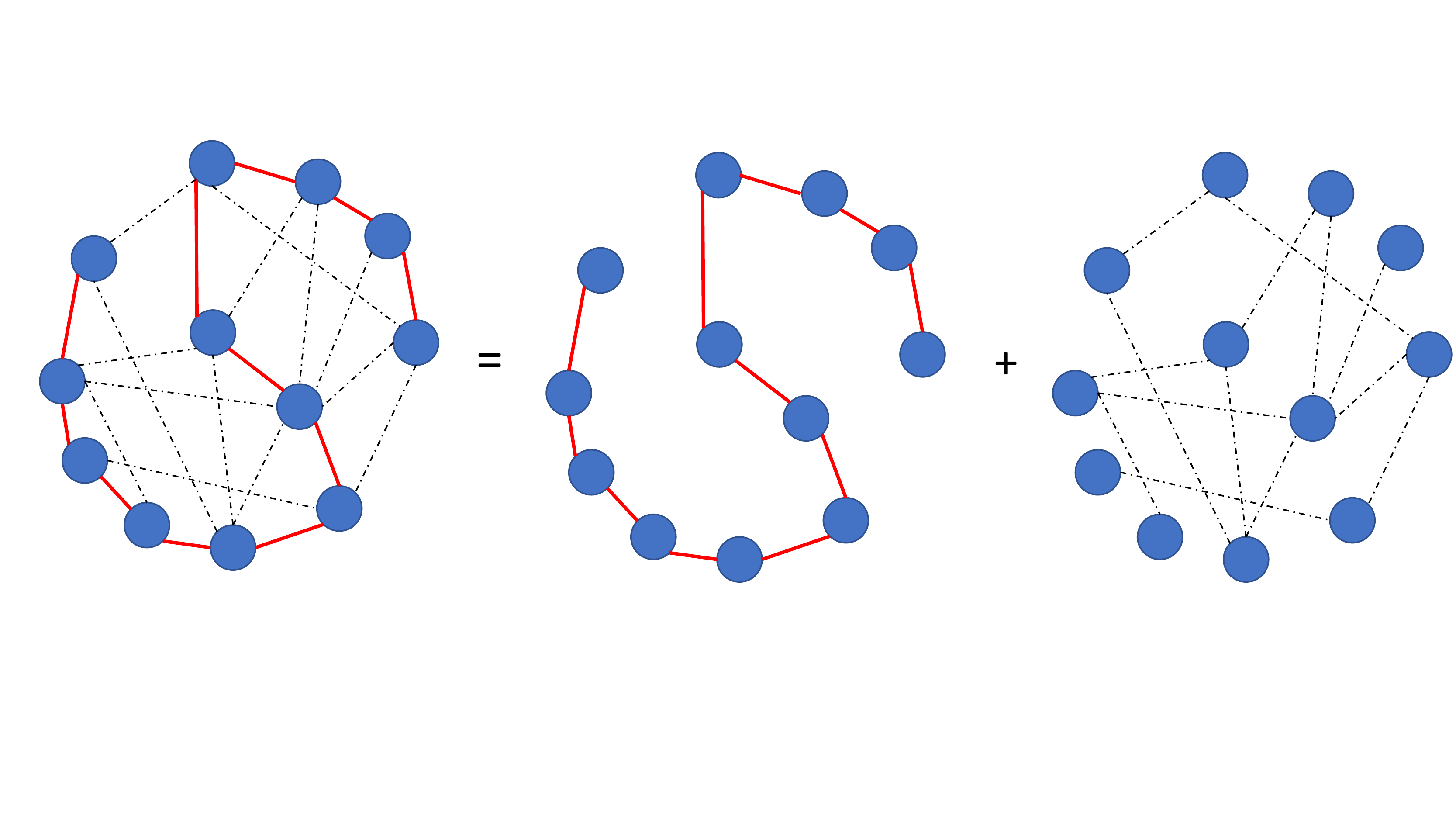}
 \caption{Protein representation in our model. The solid red and the dashed black lines represents covalent backbone and non-sequential residue bonds, which connect residues (blue spheres). The entire structure is made up by the sum of this two contributions }
 \label{fig:Pic}
 \end{figure}

 Where $\gamma_{back}$ is an arbitrary value. This  particular  choice of   $\gamma_{ij}$  comes from the idea of building a protein  potential energy as a sum of terms related to the backbone and others connected the non-bonded contacts (See Fig. \ref{fig:Pic}). 
 $\mathcal{U}$ can, indeed, be split into two parts:
 $$
 \mathcal{U} = \underbrace{\sum_{i,j=i \pm 1} \mathcal{U}_{ij}}_{\mathcal{U}_{back}}+ \underbrace{\sum_{i,j\neq i \pm 1} \mathcal{U}_{ij}}_{\mathcal{U}_{con}}
 $$
 
\begin{figure}
  \includegraphics[scale=0.5]{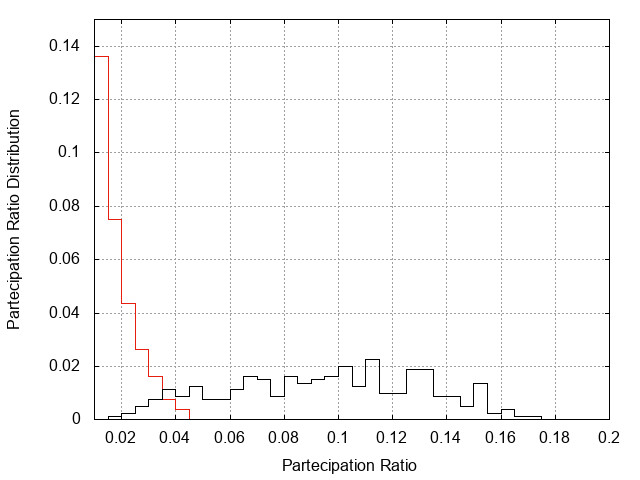}
 \caption{Participation Ratio Distribution  vs the participation ratio for $\mathcal{D}_{back}$ (red)  and $\mathcal{D}_{con}$ (black)  }
 \label{fig:Pr}
 \end{figure}

Where all the elements $\mathcal{U}_{ii}$ are $0$. Therefore, the Hessian matrix ( $\mathcal{H} $), calculated on the equilibrium positions ($\mathbf{r}_{ij}^{o}$)  can be written as  a sum of  contributions coming from the backbone  ($\mathcal{H}_{back}$) and the non-bonded  contacts ($\mathcal{H}_{con}$),
 $$
 \mathcal{H}= \nabla \nabla \mathcal{U}  =\nabla \nabla \mathcal{U}_{back}+\nabla \nabla \mathcal{U}_{con}  = \mathcal{H}_{back} +  \mathcal{H}_{con} 
 $$
Therefore, the dynamical matrices both for the backbone ($\mathcal{D}_{back}$) and the non-bonded contacts ($\mathcal{D}_{con}$) have been calculated as $ \mathcal{M}^{-1/2} \; \mathcal{H} \;\mathcal{M}^{-1/2}$ . Where $\mathcal{M}$ s  a diagonal matrix whose diagonal entries are the masses of the relative degree of freedom. We can now study separately the localization property of $\mathcal{D}_{back}$ and  $\mathcal{D}_{con}$ eigenvectors.
We examined the distribution of the participation Ratio ($P_{n}$) defined as \cite{edwards}:

$$
P_{n} =\dfrac{1}{N}(\sum_{i}^{N}|e_{n}^{i}|^{4})^{-1}
$$ 
Where $N$ is the total degrees of freedom of the system and  $e_{n}^{i}$ the $i$-th component of the $n$-th eigenvector. $P_{n}$ is usually chosen for studying localization properties since it assumes well-distinguishable values in case of  (de)localized eigenvectors. Indeed $P_{n} \sim \dfrac{1}{N}$ in case of totally localized eigenvectors or $1$ for  delocalized ones. Since we are not interested in the dependence on the eigenvalues $n$, we have calculated the distribution  ($\mathcal{P} (P_{n})$ ) of its values (Fig. \ref{fig:Pr} ).
$\mathcal{P}({P}_{l})$ shows two relevant features:
 
 (i)   Both  $\mathcal{D}_{back}$ and $\mathcal{D}_{con}$ eigenvectors turn out to be fairly localized, namely the largest values of  $P_{n}$ are below $20\%$ and $5\%$  for $\mathcal{D}_{con}$ and $\mathcal{D}_{back}$ respectively. This means they comprise a small number of degrees of freedom and they  cannot carry  vibrational energy by themselves. This implies the $E_{vib}xc$ process in proteins should involve, similarly to a disordered solid, anharmonic processes\cite{wingert}.
  
  (ii)  A not negligible discrepancy between the two sets of eigenvectors is present. The most extended one among  $\mathcal{D}_{back}$'s is  about four time shorter than the $\mathcal{D}_{con}$ one. The difference in  $E_{vib}xc$  process can be  traced back to this finding, which agrees with the experimental results since it suggests $\mathcal{D}_{con}$ are more prone to exchange vibrational energy.

Eigenvectors localization, already predicted for percolative systems and fractals \cite{gefen}, implies  an exponentially decay of their absolute values with the distance $r$, $|\mathbf{{e}}_{n}| \sim exp(-{r})$ \cite{allen}. It has been previuosly theorized that  for above mentioned systems the decay is not simply exponential, instead $|\mathbf{{e}}_{n}| \sim exp(-{r}^{d})$\cite{levy}\cite{aharony}\cite{mosco} \cite{nakayama}. We will show proteins share the same feature.  The exponent $d$ will be defined later explaining its connection with the topology. However, passing we can disclose that $d$ assumes different values depending whether we are considering the backbone or the contacts, producing different localization properties as noticed in (ii).
In order to show the presence of this decay here we reprise a theoretical picture already presented \citep{aharony} in a slightly different fashion, testing the assumptions made against  protein system.

\begin{figure}
 \includegraphics[scale=0.5]{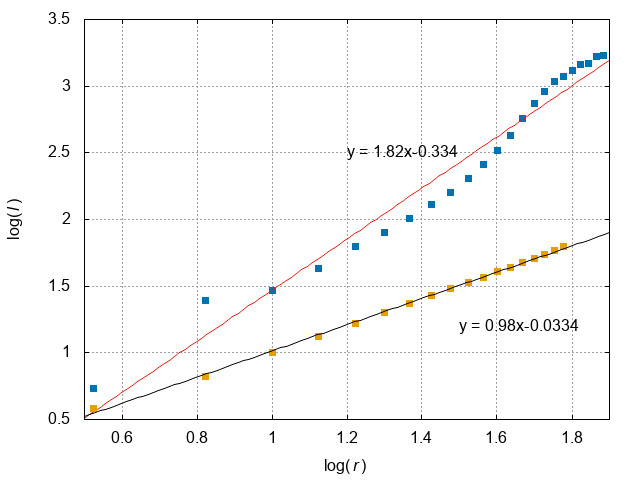}
 \caption{ $log(\ell)$ against $log(r)$ for the backbone (blue squares) and contacts (yellow squares). Continuous lines, red for the backbone and black for the contacts, are linear functions resulting from the fitting. The oscillation around the fitted value mostly present in the backbone case are $logarithmic$ oscillation already observed and predicted by the theory \cite{dewey} }
 \label{fig:l_vs_r}
 \end{figure}

The starting point are the equations of motion, which could be recast as:    
\begin{eqnarray}
-\omega_{n}^{2} \; \mathbf{e}_{n} = - \mathcal{D} \cdot \mathbf{e}_n
\end{eqnarray}
Where $\omega_{n}$ is the $n$-th eigenvalue of the dynamical matrix  $ \mathcal{D}$, which physically represents the frequency of the $n$-th eigenvector ($\mathbf{e}_{n}$). $ \mathcal{D}$   is intended to be indifferently $ \mathcal{D}_{back}$ or $ \mathcal{D}_{cont}$  as the derivation is general and comprise both cases. The difference between the two systems will be introduced later. $ \mathcal{D}$ can be thought as sum of the contribution of a dynamical matrix related to a perfect ordered system ( $ \mathcal{D}_{o}$) , whose all eigenvectors are completely extended,  and a matrix  ($\Delta$) which is the difference between $ \mathcal{D}$ and $ \mathcal{D}_{o}$. $\Delta$ includes all the contributions needed to turn a perfectly ordered system into a disordered one as a real protein. $ \mathcal{D}_{o}$  can be written as:
\begin{eqnarray}
\mathcal{D}_{o} = \Gamma \; \delta
\end{eqnarray}
Where $\Gamma $ is the Laplacian matrix associated to a d-dimensional square lattice and  $\delta$ is an arbitrary coupling constant between different degrees of freedom. It is noteworthy they are related to two different aspects of the disorder in protein systems. $\Gamma$ is linked to the topology itself. It encodes  an ordered one, whereas a protein is featured by a more complex and disordered topology.  $\delta$ is related to the variability of interactions that can be "source of  disorder" also in a topological ordered system. We assumed as a value for the latter the average value evaluated over all the $ \mathcal{D}$  entries . This appears a  physically reasonable value since  $\Delta$ can be actually thought as  "deviation" from an average coupling constant value from one hand and  an topological ordered  system on the other.
Starting from (1)  we now have:
\begin{eqnarray}
 \big[\mathcal{D}_{o} -\omega_{n}^{2}  \big] \cdot  \mathbf{e}_{n} = - \Delta \cdot \mathbf{e}_n \\ 
 \nonumber\\
 \mathbf{e}_{n} = \Big[ \mathcal{I}- \dfrac{\mathcal{D}_{o} }{\omega_{n}^{2}}  \Big]^{-1} \cdot \dfrac{\Delta}{\omega_{n}^{2}} \cdot \mathbf{e}_{n} 
\end{eqnarray}    
With $\mathcal{I}$ the identity matirx. We can now define:
 \begin{eqnarray}
 G(\omega_{n})=\Big[ \mathcal{I}- \dfrac{\mathcal{D}_{o} }{\omega_{n}^{2}}  \Big]^{-1} 
 \end{eqnarray}
In case :
 \begin{eqnarray}
 \dfrac{||\mathcal{D}_{o} ||}{\omega_{n}^{2}}  < 1  \nonumber
 \end{eqnarray}
 or equivalently
 \begin{eqnarray}
 ||\mathcal{D}_{o} ||<{\omega_{n}^{2}}    
 \end{eqnarray}
would be verified (5) can be recast as Neumann series:
 \begin{eqnarray}
 G(\omega_{n})=\sum_{p=0}^{\infty} \Big(\dfrac{\mathcal{D}_{o} }{\omega_{n}^{2}} \Big)^{p}
 \end{eqnarray}
Verify (6) requires calculating $\delta$, since $||\mathcal{D}_{o} || \sim \delta$,  and  $\omega$  for every protein system under study, which turns out to be tricky to accomplish mostly because of  $\delta$  experimental measure, which should be performed for every case. However, employing previous experimental findings, we can  estimate $\delta $ as $\dfrac{<k>}{<m>} $. Where  $<k>$ and$<m> $ are the elastic constant coupling to degrees of freedom and the mass of a residue averaged over the whole protein. Previous works on bacteriorhodopsin \cite{rico} showed  the former is $\sim 10^{-1}\; N/m$.  $<m>$ can be estimated dividing the average protein molecular weight in the Eukaryotic proteomic ($\sim 10^{1} $ KDa ) by the average length  ($10^{2}$ amino acids), obtaining $\sim 1.6  \cdot10 ^{-1}$ Kg. Therefore $\sqrt{\delta } \sim 0.7$ Thz which is  comparable with lowest vibrational frequency experimentally measured in proteins. According to this assessment, hence, (6) can be  considered approximately satisfied for real proteins. 
 It can be shown the $i,j$-th entry of $G$, as recasted in (5),  can be approximated as \cite{aharony}:

  \begin{eqnarray}
 G(\omega_{n})_{ij} \sim \Big(\dfrac{\delta }{\omega_{n}^{2}} \Big)^{\mathcal{N}_{ij}}
 \end{eqnarray}
  Where $\mathcal{N}_{ij}$ is the minimum number of steps required for connecting the $i$-th and the $j$-th over the d-dimensional square lattice. Inserting (8) in (4) and passing to the scalar equation one gets:
  \begin{eqnarray} 
e_{n}^{i} \sim \sum_{j}\sum_{k}\Big(\dfrac{\delta }{\omega_{n}^{2}} \Big)^{\mathcal{N}_{ik}}\;\dfrac{\Delta_{kj}}{\omega^{2}_{n}} \; e_{n}^{j}
 \end{eqnarray}
Thus $e_{n}^{i}$ is a sum of $N$-1 elements such that:
\begin{eqnarray} 
e_{n}^{i,(j)} \sim \sum_{k} \Big(\dfrac{\delta }{\omega_{n}^{2}} \Big)^{\mathcal{N}_{ik}}\;\dfrac{\Delta_{kj}}{\omega^{2}_{n}} e_{n}^{j}=\sum_{k} \Big(\dfrac{\delta }{\omega_{n}^{2}} \Big)^{\mathcal{N}_{ik}+1}\:\dfrac{\Delta_{kj}}{\delta} e_{n}^{j}
 \end{eqnarray}
 Where $e_{n}^{i,(j)} $ is the $j$-th elements of the summation and $k$ runs over the nearest neighbours of $j$-th degree of freedom. The  minimum values of $\mathcal{N}_{ik}+1$ obviuosly  corresponds to 
   $\mathcal{N}_{ij}$, namely the minimum number of steps connecting the $i$-th and the $j$-th degree of freedom. The summation is, hence dominated by $\Big(\dfrac{\delta }{\omega_{n}^{2}} \Big)^{\mathcal{N}_{ij}}$ 
   %if:
%   \begin{eqnarray}   
%  \Big|  \Big(\dfrac{\delta }{\omega_{n}^{2}} \Big)^{\mathcal{N}_{ij}} \; \Delta_{kj} \Big|   >> \Big|   \Big(\dfrac{\delta }{\omega_{n}^{2}} \Big)^{\mathcal{N}_{ik^{'}}+1} \; \Delta_{k^{'}j}\Big|   
 % \end{eqnarray} 
%Where $\mathcal{N}_{ik^{'}}+1$ is the minimum value larger then ${\mathcal{N}_{ij}}$ among $j$'s nearest neighbours. In the worst case   $\mathcal{N}_{ik^{'}}+1-\mathcal{N}_{ij}=1$ thus (11) becomes:
%\begin{eqnarray}   
 %\Big|  \dfrac{\Delta_{kj}}{\Delta_{k^{'}j} } \Big| >>    \dfrac{\delta }{\omega_{n}^{2}}  
  %\end{eqnarray} 
%Assuming the right hand side of (12) $\sim 1$, since $\Delta_{k(k^{'}),j} $ should be on average the same order of magnitude, this condition is similar to (6). 
and (10) becomes:
\begin{eqnarray} 
e_{n}^{i,(j)} \sim \Big(\dfrac{\delta }{\omega_{n}^{2}} \Big)^{\mathcal{N}_{ij}} e_{n}^{j}
 \end{eqnarray}
 It is evident from (13) the connection between two differnt degree of freedom occurs applying $\Big(\dfrac{\delta }{\omega_{n}^{2}} \Big)$  $\mathcal{\tilde{N}}$ times  , where $\mathcal{\tilde{N}}$  is the minimum number of steps connecting them. Therefore (13) holds for all couples of degree of freedom sharing the same minimum number of connecting steps, regardless the kind of topology, and  can be re-written taking into accounts only this variable.
 \begin{eqnarray} 
e_{n}^{\mathcal{\tilde{N}}} \sim \Big(\dfrac{\delta }{\omega_{n}^{2}} \Big)^{\mathcal{\tilde{N}}} e_{n}^{0}
 \end{eqnarray}
 Where $e_{n}^{\mathcal{\tilde{N}}}$ and $e_{n}^{0} $  are degree of freedoms separated by $\mathcal{\tilde{N}}$  and $0$ number of steps respectively by  the  degree of freedom $j$, which corresponds to the largest absolute value among  $\mathbf{e}_n$ entries that we set as the "orgin".
 It is now feasible define a localization over the minimum number of steps  $\Lambda(\omega_{n})$ as :
 \begin{eqnarray} 
-\dfrac{1}{\Lambda(\omega_{n})}=\lim_{\mathcal{\tilde{N}} \rightarrow \infty} \dfrac{1}{\mathcal{\tilde{N}}} \log{\Big|\dfrac{\delta }{\omega_{n}^{2}} \Big|^{\mathcal{\tilde{N}}}}=\log{\Big|\dfrac{\delta }{\omega_{n}^{2}} \Big|}
 \end{eqnarray}
 thus: 
  \begin{eqnarray} 
|e_{n}^{\mathcal{\tilde{N}}}| \sim exp \Big(-\dfrac{\mathcal{\tilde{N}}}{\Lambda(\omega_{n})} \Big) 
 \end{eqnarray}
Dividing and multiplying the argument of the exponential for the average  eucleadian distance associated to one step on the structure one gets:
  \begin{eqnarray} 
|e_{n}| \sim exp \Big(-\dfrac{\mathcal{\ell}}{\Lambda^{*}(\omega_{n})} \Big) 
 \end{eqnarray} 
Where  $\ell$ is the  minimum average distance  between two different degree of freedom, also called $chemical \: dimension$, and   $\Lambda^{*}(\omega)$ is the corresponding average localization length. 
Eq. (17) tells us eigenvectors decay exponentially along the path of minimum distance, which  can account for  (i). Explaining (ii), instead, requires studying  the relation between $\ell$  and the Euclidean distance $r$.
In case of spatially homogeneous solids $\ell = r$, however when one deals with inhomogeneous structures as percolative system or fractals it has been 
empirically shown  the relation becomes a power law  as  $\ell \sim r^{d_{min}}$  \cite{herrmann}. Previous experimental studies showed protein can present, 
in a statistical sense \cite{banerji}, properties featuring fractals. This similarity  consist in power law  scaling regarding particular quantities as, for instance: 
The radius of gyration,\cite{dewey},  the protein mass (as well  as the density) within a sphere \cite{dewey}  \cite{leitner} or  the surface "roughness"  \cite{lewis}  \cite{leitner} 
, furthermore inderect experimental evidences suggested a scaling of the density of collective motions with the frequency typical of 
"fractons" \cite{stapleton} \cite{helman} \cite{alexander}. Therefore, it is reasonable that  $\ell$ can show  a power law scaling as well. Obviously the scaling law, if 
present, should be different whether we consider only  the backbone  or the contacts since, according to definition of $\ell$, the connections between degree 
of freedoms (residues) are different in the two cases. This idea has been verified numerically obtaining a value of $d_{min}$  $\sim$ 1.8 and $\sim$ 1 for the 
backbone and the contacts respectively (See. fig.\ref{fig:l_vs_r} ).

\begin{figure}
 \includegraphics[scale=0.5]{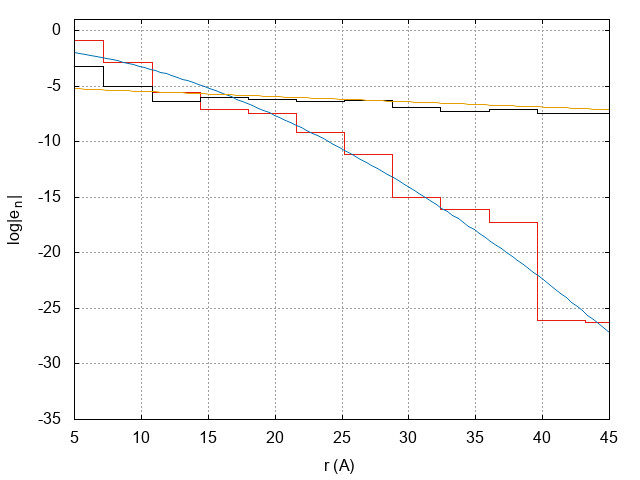}
 \caption{$log|e_{n}|$ for $\mathcal{D}_{back}$ (red)  and $\mathcal{D}_{con}$ (black) taken as an example among one of the investigated protein system (PDBID: 1hty). Continuous lines have been drawn as guide to the eye. Light blue line is  proportional to $-x^{1.8}$ , the yellow one  to $-x$. }
 \label{fig:Psi}
 \end{figure}

The backbone presents a larger $d_{min}$ showing a higher degree of inhomogeneity, differently the structure made up by contacts resembles a more homogeneous one thanks to the large number of interactions present. Therefore, according to (17),  $\mathcal{D}_{back}$ eigenvectors decay with $r$ as  $\sim exp(-r^{1.8})$.This particular decay has been observed before and the resulting localization has been  named  $superlocalization$\cite{levy}\cite{nakayama}. The $\mathcal{D}_{back}$ eigenvectors, representing collective motions taking place only on the backbone are $superclocalized$  and  less prone to exchange vibrational energy among  distant residues than the collective motions over non-adjacent residue contacts that are simply exponentially localized (see. Fig \ref{fig:Psi}). 
In conclusion we have explained why  $E_{vib} \; xc$ occurs mainly through weaker residue-residue contacts instead of backbone bonds. This experimental finding, albeit   at first glance counterintuitive, agrees with our picture. The exponent $d_{min }$  which governs the collective motions exponential  decay with the Euclidean distance is different whether we consider the backbone  ($d_{min} \sim 1.8$) or  non-adjacent residue-residue interactions  ($d_{min} \sim 1$). This implies a $superlocalization$ of collective motions associated to the backbone.

\bibliographystyle{unsrt}
\bibliography{article_draft}

\end{document}